# PHASE-FIELD MODELING OF SOLIDIFICATION IN LIGHT-METAL MATRIX NANOCOMPOSITES

Tamás Pusztai[1], László Rátkai[1], Attila Szállás[1], László Gránásy[1,2]

[1]Wigner Research Centre for Physics, H-1525 Budapest, POB 49, Hungary

[2]BCAST, Brunel University, Uxbridge, Middlesex, UB8 3PH, United Kingdom

Keywords: Metal Matrix Nanocomposites, Solidification, Phase-field modeling, Particle pushing

## Abstract

The quantitative phase-field approach has been adapted to model solidification in the presence of Metal Matrix Nanocomposites (MMNCs) in a single-component liquid. Nanoparticles of fixed size and shape are represented by additional fields. The corresponding equations of motion are assumed to ensure relaxation dynamics, and can be supplemented by random forces (realizing Brownian motion) or external fields. The nanoparticles are characterized by two model parameters: their mobility and the contact angle they realize with the solid-liquid interface. We investigate the question how grain size distribution can be influenced by heterogeneous nucleation on the nanoparticles and by the front-particle interaction. We explore, furthermore, how materials and process parameters, such as temperature, density and size/shape distribution of the nanoparticles, influence microstructure evolution.

## Introduction

The Al and Mg based MMNCs are promising candidates for automotive, aerospace, and defense applications due to the drastic weight savings and exceptional properties.[1,2] The distribution of solid nanoreinforcers (10 to 500 nm) in molten alloys is influenced by various phenomena, such as coagulation of the nanoparticles, the interaction between the particles and propagating solidification fronts (particle pushing and engulfment), whereas the nanoparticles themselves may act as heterogeneous nucleation sites for solidification, to mention a few. Owing to the complexity of the solidification process, development of mathematical models that capture the main effects is desirable. Phase-field (PF) modeling is one of the most potent tools for describing microstructure evolution during solidification. The PF approach has already been used to address homogeneous[3] and heterogeneous[4,5] crystal nucleation, dendritic growth,[3,6,7] polycrystalline freezing,[2,8-11] and particle pushing,[12] etc., however, a PF model that addresses all these processes simultaneously is yet unavailable. (A similar model is expected to be useful in modeling the effect of grain refiners, however, on a different size scale.) Herein, we make a step towards such a unified model, and present a PF approach that incorporates heterogeneous nucleation via appropriate boundary conditions and noise representing the thermal fluctuations in addition to a simplified treatment of the particle-front interaction. Illustrative computations are then presented in two dimensions (2D), which explore the effect of nanoparticles on the solidification of pure Mg.

## The Applied Phase-Field Models

In order to predict *homogeneous nucleation* in undercooled Mg (which results will serve as reference for the subsequent simulations), we use a detailed PF model of homogeneous nucleation that has achieved reasonable agreement with molecular dynamics simulations for the Lennard-Jones system and with experiments for the ice-water system without adjustable parameters.[3] Here, the Euler-Lagrange equation is solved with appropriate boundary conditions to determine the properties of the homogeneous nucleus, which is then used to compute the nucleation rate as described in Reference 3. The nucleation time is computed as follows: $\tau_1 = 1/(VJ_{SS})$, where $V = L^3$ is the control volume, $L$ the linear size of the simulations, whereas $J_{SS}$ is the steady-state nucleation rate (number of nuclei formed in unit time and volume).

Next, we present the PF model for *heterogeneous nucleation and particle-front interaction*. For the sake of simplicity, we present here the version that handles a single foreign particle. In previous studies, surfaces of the foreign particles have been represented by boundary conditions.[4,5] Physically, this corresponds to sharp interfaces between the simulation volume and the foreign particle (wall) region. However, to model particle pushing, this approach is generally inappropriate, since it is incompatible with a sub-pixel translation of the foreign surfaces. To overcome this limitation, we have generalized Models A of Reference 5 to allow for *diffuse walls*, i.e., where the interface between the impurity particle and simulation volume is not mathematically sharp. This has been achieved by reformulating our phase-field model so that the boundary conditions defining the foreign particles are converted to a volumetric contribution via introducing a new phase field $\phi_w(r)$ for the particle[5]

$$F = \int dV \left\{ Z(\phi)|\nabla \phi_w| + \left[ f(\phi) + \frac{\varepsilon^2}{2}(\nabla \phi)^2 \right](1-\phi_w) \right\}$$

as opposed with the original form[4,5,13]

$$F = \int dA \, Z(\phi) + \int dV \left[ f(\phi) + \frac{\varepsilon^2}{2}(\nabla \phi)^2 \right]$$

where $\phi$ is the solid-liquid order parameter (0 in the liquid and 1 in the solid), $Z(\phi)$ is the surface function of Model A that determines the wetting properties of the surfaces, $\phi_w(r)$ is the continuous "wall field" that is 1 inside the volume occupied by the impurity particles and 0 outside with a monotonic transition in a narrow region of width $\delta$ between, $f(\phi)$ and $\frac{1}{2}\varepsilon^2(\nabla \phi)^2$ are the usual free energy density and square gradient terms of phase field theory. It is easy to see, that in the $\delta \to 0$ limit, the generalized functional falls back to the original functional, therefore the results obtained with the diffuse wall approach expected to be in good agreement with the results of the original sharp wall approach.

The time evolution of the phase field $\phi(r, t)$, which monitors solidification is determined by the following equation of motion

$$\dot{\phi} = -M_\phi \left\{ \left[ \frac{\partial f}{\partial \phi} - \varepsilon^2 \nabla^2 \phi \right] (1 - \phi_W) + \left[ \frac{dZ}{d\phi} + \varepsilon^2 (\nabla \phi \cdot \mathbf{n}) \right] |\nabla \phi_W| \right\}$$

where $M_\phi$ is the phase-field mobility, associated with the translational diffusion coefficient, and $\mathbf{n}$ is the surface normal of the foreign particle.

To address the dynamics of particle pushing, we derive an equation of motion for the particle in 2D. Since the circular particle is assumed to be of fixed shape and size, it can be described mathematically by a simple radial function as

$$\phi_W(x, y) = \varphi_w \left( \sqrt{(x - x_0)^2 + (y - y_0)^2} \right).$$

(In practice, a *tanh* profile of width $\delta$ in the order of 1 nm is used.) This way, all the dependence of the free energy of the system on $\phi_w$ can be expressed with the $x_0$ and $y_0$ central coordinates, and the equations of motion for this coordinates is assumed to follow relaxational dynamics in the form of

$$v_x = -M_w \frac{\partial F}{\partial x_0} = -M_w \int dV \left\{ \begin{bmatrix} \left[ f(\phi) + \frac{\varepsilon^2}{2} (\nabla \phi)^2 \right] \varphi'_w \\ -Z(\phi) sign(\varphi'_w) \varphi''_w \end{bmatrix} \frac{x - x_0}{r} \right\},$$

where $v_x$ is the velocity of the impurity particle and $M_w$ is the mobility of the particle that can be related to viscosity and particle size. Similar equation holds for the $y$ component.

Generalization of this approach to several foreign particles is straightforward: for $n$ particles, one needs to introduce $n$ particle fields, $\phi_{w,i}(r)$ where $i = 1, …, n$. This also means that $n$ equations of motion, similar to the one above, have to be introduced to describe the time evolution of the particle fields. In addition, a hard-sphere interaction is assumed between the foreign particles.

**Materials Properties and Model Parameters**

The thermal properties of Mg have been taken from the SGTE database.[14] In computing the heterogeneous nucleation rate PF model parameters have been fixed so that the free energy of the equilibrium solid-liquid (89.9 mJ/m$^2$) has been taken from molecular dynamics simulations,[15] whereas 10%–90% the interface thickness has been assumed to be 2 nm. We have "calibrated" the two 2D simulations so that in the $L^2$ simulations the first homogeneous nucleation event happens at the same time as in an $L^3$ volume as emerges from the Euler-Lagrange computations. This has been achieved by introducing (i) a temperature dependent effective solid-liquid interface free energy: $\gamma = (-1.4162 \times 10^{-1} + 3.2272 \times 10^{-4} T)$ J/m$^2$, and (ii) the phase field mobility has been set to $M_\phi = 0.16$ m$^3$/Js. In the simulations, unless stated otherwise, $L = 512 \times \Delta x = 0.128$ μm, as $\Delta x = 0.25$ nm, whereas the time step has been $\Delta t = 10^{-10}$ s. The system has been coupled to a 700 K block via a heat transfer coefficient of $\lambda = 5 \times 10^4$ W/(m$^2$K), and the heat release during crystallization is also taken into account (distributed uniformly in the simulation box). Other data used here are the melting point $T_f = 923.0$ K, the heat of fusion $\Delta H_f = 8.48$ kJ/mol, the specific heat $C_p = 24.87$ J/mol/K, and the molar volume $V_m = 1.398 \times 10^{-5}$ m$^3$/mol. $M_W = 8.0 \times 10^{-12}$ m$^3$/Js. We assume that the 2D simulation refers to a thin layer of $h = 10\Delta x$ thickness. Accordingly, we used in the simulations a fluctuation-dissipation noise of amplitude $\zeta = [2k_B T M_\phi / (\Delta t \Delta x^2 h)]^{-1/2}$. Some of the tests performed in validation of the model have been performed with the properties of nickel. Unless not stated explicitly otherwise, the results refer to Mg.

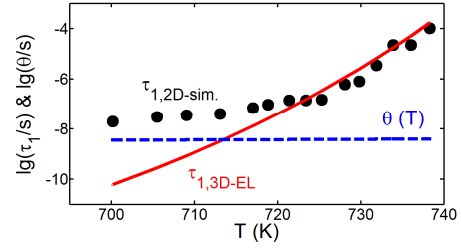

Figure 1. Comparison of nucleation times from 2D simulations (black circles) for homogeneous nucleation performed with adjusted interface free energy and phase-field mobility with nucleation times from steady-state nucleation rate from Euler-Lagrange computations (red line). For comparison the estimated incubation time $\theta$ is also shown, indicating that the deviation at lower temperatures is probably due to the nucleation transient. Note the reasonable agreement above 720 K.

**Results and Discussion**

Calibration

Calibration of the 2D simulations in the case of homogeneous nucleation has been done so that the first nucleation time coincides in the $L^2$ simulation with that from the detailed Euler-Lagrange computations, as shown in Figure 1. When they become available molecular dynamics or experimental data can also be used in calibrating the simulations.

Validation of the Model

Testing the Equilibrium Contact Angle. We have switched off cooling, but retained heat release during solidification, which heats up the simulation box to an equilibrium temperature. With appropriate seeding the process can be accelerated. Results for flat interfaces of contact angle $\vartheta$ = 30°, 60°, 90°, and 120° are presented in Figure 2. Although a full equilibrium is not yet achieved, the contact angles are close to the prescribed values in all cases.

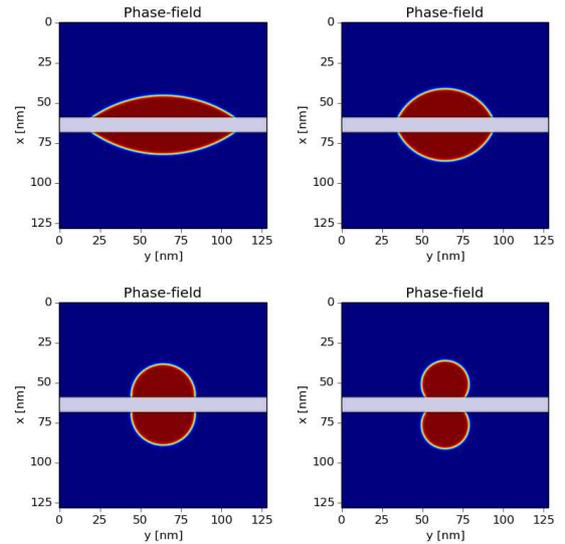

Figure 2. Crystallites approaching equilibrium when prescribing contact angles of 30°, 60°, 90°, and 120° in the PF simulation.

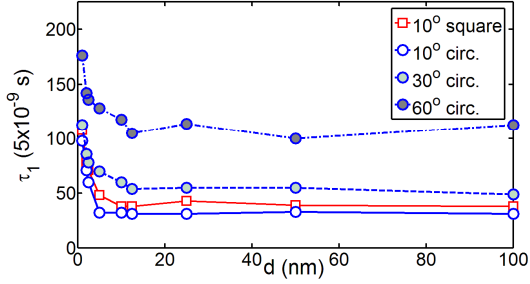

Figure 3. Heterogeneous nucleation time vs. particle size (diameter or edge length) as a function of contact angle in systems, where the total length of the particle perimeters is the same, which length is distributed to a decreasing number of particles from left to right. Circular and square particles behave similarly.

Testing of Heterogeneous Nucleation. On the basis of theoretical consideration one expects that the time of appearance of the first nucleus depends only on the total length of the particle perimeters, while to a first approximation, it is independent of the number of particles, so far as the individual particle sizes are substantially larger than the homogeneous nucleus. We have tested this for $\vartheta$ = 10°, 30°, and 60°: we have distributed the total perimeter length among an increasing number of particles of uniform size. The results are shown in Figure 3 for both circular and square-shaped particles. Indeed, for larger particles ($L > 10$ nm), we find a reasonably constant nucleation time, however, for smaller particles the nucleation efficiency decreases as expected theoretically.[16]

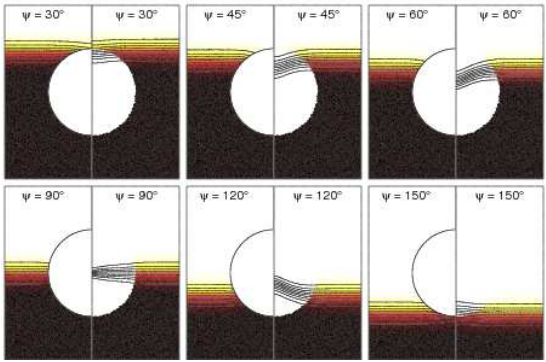

Figure 4. Stationary configurations of particles pushed by a planar solidification front in Ni at an undercooling of 10K. The panels show the half cross sections of the cylindrically symmetric 3D steady state solutions travelling with the velocity corresponding to the velocity of the free solid-liquid interface. The transition from black to white indicates the transition from the solid to liquid phase, while the contour lines corresponding to $\phi_0$ = 0.1, 0.2, ... , 0.9 are also plotted. The left and right sides of the double panels show the solutions based on the sharp and diffuse wall approaches, respectively. In the case of diffuse walls, the solution assigns $\phi$ values to the interior of the particle, which, for better visibility, is shown only by its contour lines and not by the respective colors. The diameter of the particle is 40nm, while the size of the simulation box is 40×80 nm in all panels.

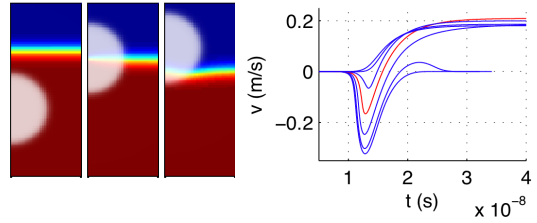

Figure 5. Particle pushing by a planar solidification front as predicted by the phase-field theory in Ni. The left three panels correspond to $\vartheta$ = 30°, 90°, and 150° and show the snapshots of the radial cross sections of the cylindrically symmetric 3D solutions approximately 2×10$^{-8}$ s after the solidification front reached the particle. (The central snapshot corresponds to the right side of the 90° panel in Fig. 4.) The rightmost panel shows the velocity profiles evaluated from the simulations. After a transient period, the particle is either engulfed by the moving solid-liquid interface (at low contact angles) or pushed (at high contact angles). At the velocity minimum ($t \approx 1.3 \times 10^{-8}$ s), the curves from bottom to top correspond to contact angles $\vartheta$ = 0°, 30°, 60°, 90°, 120°, 150°, and 180°, respectively.

Testing of Particle-Front Interaction (Steady State). Using the properties of Ni, we have determined the steady state configurations where the particle has the same speed as the propagating solidification front. This configuration can be found as the time independent solution of the equation of motion expressed in a local coordinate system moving with the front. We utilized the cylindrical symmetry of the problem, reducing thus the dimensionality of the solution from 3 to 2. Since in the time independent solutions no displacement of the impurity particle is required (the particle moves together with the frame), both the sharp and the diffuse wall approaches can be used to address this problem, making this scenario an ideal candidate for comparing the results obtained by sharp and diffuse interfaces. As displayed by Figure 4, the two solutions are almost identical outside the impurity particle, i.e., in the physically relevant domain.

Testing of Particle-Front Interaction (Dynamics). Using the properties of pure Ni, we have solved numerically the equation of motions for the phase field and the particle fields. We have adjusted $M_w$ so that velocity of the particle coincides with the velocity of the steady-state solution for a particle with 90° contact angle at the solid-liquid interface. Then we varied the contact angle between 0 and 180° (see Figure 5). We have found that below a critical contact angle that falls between 30° and 60°, the particle is engulfed, while above this critical contact angle the particle is pushed by the advancing solidification front.

Size Distribution of Particles. We have performed simulations for 200 circular particles of average radius of 10$\Delta x$ and standard deviation of 5$\Delta x$ of two kinds of monodisperse distribution, which show uniform and normal distributions, and a bimodal size distribution. The mobility of the particles is reduced so much that particle motion is entirely negligible during the simulations. The results are summarized in Figures 6(a)-(c). Apparently, features of the size distribution of the foreign particles are reflected by the grain size distribution. For example, a bimodal particle distribution of foreign particles often leads to a bimodal grain size distribution (Fig. 6(c)). A general observation is that crystallization starts in areas, where the particle density is the largest.

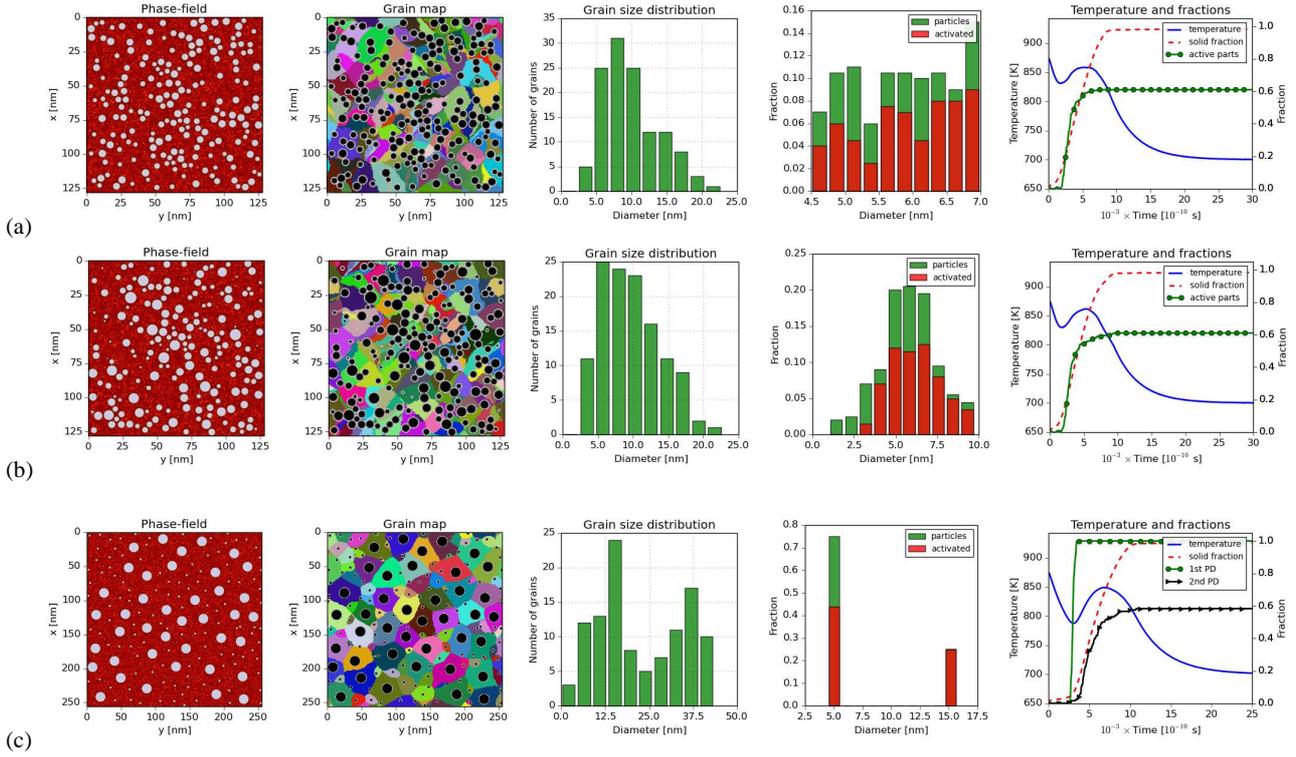

Figure 6. Solidification in the presence of foreign particles of $\vartheta = 10°$: 200 particles of (a) uniform, (b) normal, and 160 particles of (c) bimodal size distribution (40 large, 120 small). The final (fully solidified) stage is shown. From left to right, the phase-field map, the grain map (different colors correspond to different grains), the size distribution of the crystalline grains, the size distribution of the foreign particles (red fraction participated in nucleation), and the time evolution for the temperature (solid), the crystalline fraction (dashed), and the fraction of activated foreign particles (circles and triangles for the two sizes) are shown. Grid sizes: (a), (b) 512×512 and (c) 1024×1024.

Particle Motion. Finally, relying on the assumption of overdamped dynamics, we have set the particle mobility to $M_W = 1/(6\pi\mu R)$, which follows from the expression for the Stokes velocity, where $\mu = 1$ mPa.s is assumed for the dynamic viscosity in a simulation performed under conditions identical to those used for computing Figure 6(c). As a result, the particles started to display Brownian motion-like behavior due to the capillary forces emerging from the phase-field fluctuations (see Figure 7). This motion is, however, soon stopped by the growing crystals. Owing to the small contact angle (10°), the particles are easily engulfed into the growing crystal. The grain size distribution is fairly similar to the one obtained without particle motion. Note the differences in the initial and final spatial arrangement of the foreign particles, which originate from their stochastic motion. For comparison, we performed a similar simulation, however, now the contact angle of the small particles have been set to 120°. As a result, the small particles are being pushed by the solidification front (Figure 8), and accumulate at the grain boundaries and in liquid pockets. The latter often lead to the formation of particle islands, consisting of closely packed small particles (located on a hexagonal lattice). Hence, particle pushing leads to a drastic spatial rearrangement of the small particles.

Larger simulations are yet needed to reveal details of the connection between the size distributions of the foreign particles and the grain size distribution.

## Summary

We have presented a simple phase-field model that addresses crystal nucleation and growth in the presence of foreign particles characterized by a size and viscosity related mobility and a contact angle. The model has been tested concerning equilibrium and dynamical contact angles, for heterogeneous nucleation and particle pushing, and has shown a reasonable behavior. Illustrative simulations have been performed that show that the size distribution of the foreign particles is reflected in the size distribution of the crystalline particles. For example, a bimodal particle size distribution of the foreign particles may lead to a bimodal grain size distribution. It has also been demonstrated that particle motion can be taken into account, and it may influence the solidification microstructure considerably, especially if the contact angle of the particles is large. Work is underway to extend the capabilities of the model towards quantitative modeling of particle induced solidification in alloys.

## Acknowledgments

This work has been supported by in the framework of the EU FP7 Collaborative Project "EXOMET" (contract no. NMP-LA-2012-280421, co-funded by ESA).

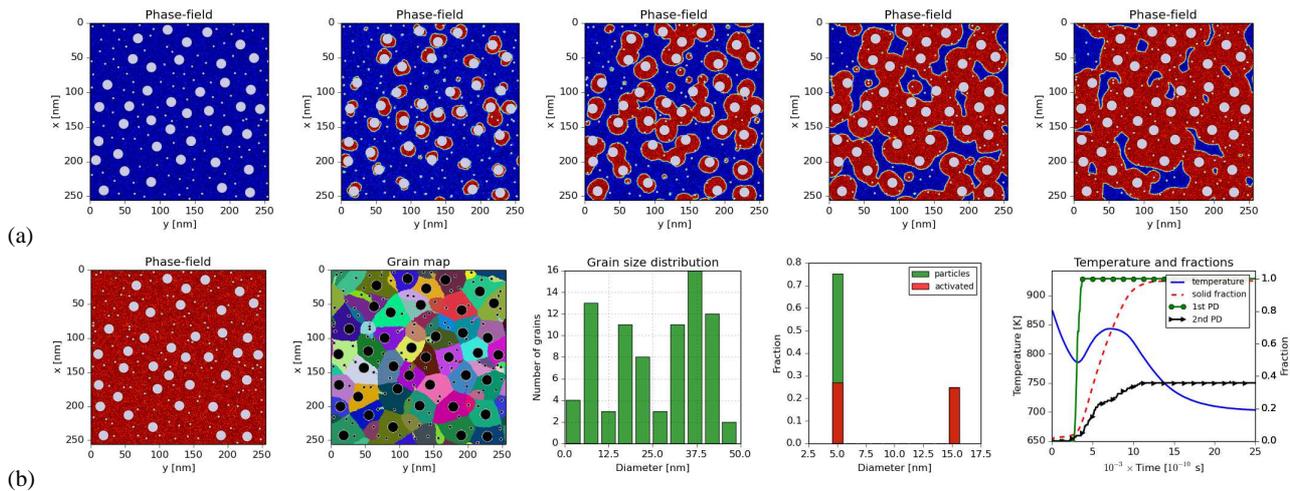

Figure 7. As the simulation shown in Figure 6(c), however, the particles *move due to capillary forces emerging from phase-field fluctuations*: (a) Snapshots showing of the phase field map. (b) The same sequence of panels as displayed in Fig. 6. Particle motion has changed somewhat the grain size distribution and the nucleation kinetics on the small particles relative to Figure 6(c).

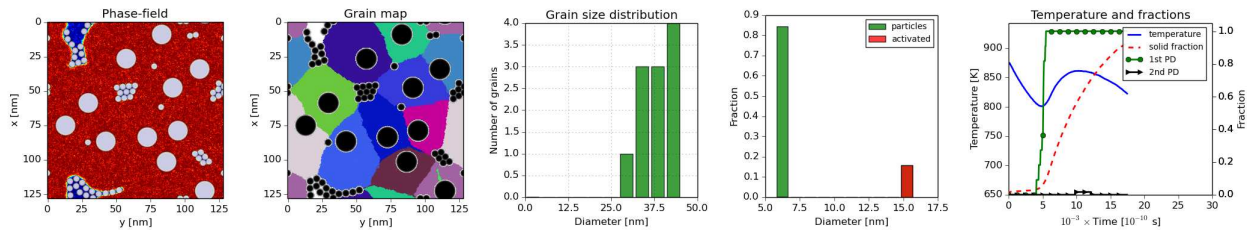

Figure 8. Late stage of freezing in the presence of 70 particles (11 large of $\vartheta = 10°$, 59 small of $\vartheta = 120°$). Note that particle pushing drives the small particles to the grain boundaries or into liquid pockets (blue and white regions in panels No. 1 and 2, respectively). In the liquid pockets hexagonal particle arrangements form eventually, as freezing progresses. Note that only the large particles induce nucleation.

**References**

1. A. Mortensen and J. Lorca, "Metall Matrix Composites," *Ann. Rev. Mater. Res*, 40 (2010), 243–270.
2. P.K. Rohatgi and B. Schultz, "Lightweight Metal Matrix Nanocomposites – Stretching the Boundaries of Metals," *Material Matters*, 2 (2007), art. no.16.
3. L. Gránásy, T. Börzsönyi, and T. Pusztai, "Nucleation and Bulk Crystallization in Binary Phase Field Theory," *Phys. Rev. Lett..*, 88 (2002), art. no. 206105.
4. L. Gránásy, T. Pusztai, D. Saylor, and J. A. Warren, "Phase Field Theory of Heterogeneous Crystal Nucleation," *Phys. Rev. Lett.*, 98 (2007) art. no. 035703.
5. J. A. Warren, T. Pusztai, L. Környei, and L. Gránásy, "Phase Field Approach to Heterogeneous Crystal Nucleation in Alloys," *Phys. Rev. B*, 79 (2009), art. no. 014204.
6. J. A. Warren and W. J. Boettinger, "Prediction of Dendritic Growth and Microsegregation Patterns in a Binary Alloy Using the Phase-Field Method," *Acta Metall. Mater.*, 43 (1995), 689–703.
7. A. Karma and W.-J. Rappel, "Quantitative Phase-Field Modeling of Dendritic Growth in Two and Three Dimensions," *Phys. Rev. E*, 57 (1998), 4323–4349.
8. R. Kobayashi, J. A. Warren, and W. C. Carter, "Vector Valued Phase Field Model for Crystallization and Grain Boundary Formation," *Physica D*, 119 (1998), 415–423.
9. L. Gránásy, T. Pusztai, J. A. Warren, J. F. Douglas, T. Börzsönyi, and V. Ferreiro, "Growth of 'dizzy dendrites' in a random field of foreign particles," *Nature Mater.*, 2 (2003), 92–96.
10. L. Gránásy, T. Pusztai, G. Tegze, J. A. Warren, and J. F. Douglas, "Growth and Form of Spherulites" *Phys. Rev. E*, 72 (2005), art. no. 011605.
11. H. Henry, J. Mellenthin, and M. Plapp, "Orientation-field Model for Polycrystalline Solidification with a Singular Coupling between Order and Orientation" *Phys. Rev. B*, 86 (2012), art. no. 054117.
12. M. Ode, J. S. Lee, S. G. Kim, W. T. Kim, and T. Suzuki, "Numerical Simulation of the Critical Velocity for Particle Pushing/Engulfment Transition in Fe–C alloys Using a Phase-Field Model" *ISIJ Int.*, 40 (2000), 153–160.
13. T. Pusztai, G. Tegze, G.I. Tóth, L. Környei, G. Bansel, Z. Fan, and L. Gránásy, "Phase-field Approach to Polycrystalline Solidification Including Heterogeneous and Homogeneous Nucleation" *J Phys.: Condens. Matter*, 20 (2008), art. no. 404205.
14. A. T. Dinsdale, "SGTE Data for Pure Elements," *CALPHAD* 15 (1991), 317-425.
15. Y. G. Xia, D. Y. Sun, M. Asta, and J. J. Hoyt, "Molecular Dynamics Calculations of the Crystal-Melt Interfacial Mobility for Hexagonal Close-packed Mg" *Phys, Rev. B*, 75 (2007), art. no. 012103.
16. X. Y. Liu, K. Maiwa, and K. Tsukamoto, "Heterogeneous Two-dimensional Nucleation and Growth kinetics" *J. Chem. Phys.*, 106 (1997), 1870–1879.